\documentclass[11pt,preprint,aps,amssymb]{revtex4}
\def\be{\begin{equation}}
\def\ee{\end{equation}}
\def\bea{\begin{eqnarray}}
\def\beas{\begin{eqnarray*}}
\def\eea{\end{eqnarray}}
\def\eeas{\end{eqnarray*}}
\def\ba{\begin{array}}
\def\ea{\end{array}}
\begin{document}

\begin{center}
{\Large Discrete symmetries, invisible axion and lepton number
symmetry in an economic 3-3-1 model }

\vskip .3cm \normalsize {\bf Alex G. Dias$^a$, C. A de S.
Pires$^b$  and P. S. Rodrigues da Silva$^c$} \vskip .3cm \it (a)
Instituto de F\'{\i}sica,
Universidade de S\~ao Paulo, Caixa Postal 66.318, 05315-970,\\
S\~ao Paulo - SP, Brazil.\\ \it (b) Departamento de F\'{\i}sica,
Universidade Federal da Para\'{\i}ba, Caixa Postal 5008,
58051-970, Jo\~ao Pessoa - PB, Brazil.\\
\it (c) Instituto de  F\'{\i}sica Te\'{o}rica, Universidade
Estadual Paulista, Rua Pamplona 145, 01405-900 S\~{a}o Paulo -
SP, Brazil. \vskip .3cm
\end{center}
\begin{abstract}
We show that Peccei-Quinn and lepton number symmetries  can be a
natural outcome in a 3-3-1 model with right-handed neutrinos after
imposing a $Z_{11}\otimes Z_2$ symmetry. This symmetry is
suitably accommodated in this model when we augmented its
spectrum by including merely one singlet scalar field. We work
out the breaking of the Peccei-Quinn symmetry, yielding the
axion, and study the phenomenological consequences. The main
result of this work is that the solution to the strong CP problem
can be implemented  in a natural way, implying an invisible axion
phenomenologically unconstrained, free of domain wall formation
and constituting a good candidate for the cold dark matter.
\end{abstract}

\maketitle
\section{Introduction}
\label{secint}

The Standard Model (SM) of strong and electro-weak interactions,
$SU_C(3)\otimes SU_L(2)\otimes U_Y(1)$, has shown its
extraordinary accuracy in explaining many features of particle
physics along the years. Among the issues not covered by this
successful model there is the fact that the QCD vacuum has a
nontrivial structure revealed by its non-perturbative regime,
implying the so called strong-CP or $\theta$ problem (the subject
is widely reviewed in Ref.~\cite{strongcp}). The violation of CP
by strong interactions appears in the theory after the
introduction of instanton solution to solve the $U_A(1)$
problem~\cite{thooft}. It induces the so called $\theta$-term in
the QCD Lagrangian, which violates $P$, $T$ and $CP$. Additional
electro-weak effects change this term proportionally to $Det[M]$,
where $M$ is the quark mass matrix. The effective $\theta$-term,
$\theta_{eff}$, is observable through the electric dipole moment
of neutron, whose experimental bound implies the upper limit
$\mid\theta_{eff}\mid < 10^{-9}$~\cite{PDG1}. The smallness of
$\theta_{eff}$ is what we call the strong-CP problem.

Among the several solutions proposed to solve the Strong-CP
problem, there is one which is particularly elegant. It was
introduced by Peccei and Quinn~\cite{PQ}, and consists of
imposing a global chiral symmetry, known as Peccei-Quinn (PQ)
symmetry, $U_{PQ}(1)$, to the classical Lagrangian so that the
dynamics of the theory sets $\theta_{eff}$ to zero. Due to the
breaking of PQ symmetry a massless pseudo-scalar is generated,
the axion, which couples linearly to the axial anomaly. When this
axion develops a vacuum expectation value (VEV), $v_{PQ}$, it
produces a further displacement on $\theta_{eff}$, making it
disappear  in favor of a dynamical field, the physical axion,
eliminating the strong-CP violating term of the theory. The
breaking of PQ symmetry brings a new scale into the theory,
$f_{PQ}$, the axion decay constant, bounded by astrophysical and
cosmological data, and its allowed range is, $10^9~{\mbox GeV} <
f_{PQ} < 10^{12}$~GeV~\cite{Raffelt}.

The first class of models introducing the axion via PQ symmetry in
the context of SM were obtained by Weinberg and
Wilczek~\cite{weinwil}. This axion was soon shown to be
unrealistic mainly due to its non-suppressed coupling to light
matter fields~\cite{PecceiHomma}, which happens when $v_{PQ}$ is
of order of the electro-weak scale. Viable models to solve the
strong-CP, introducing an invisible axion, were devised by Kim and
independently by Shifman, Vainshtein and Zakharov~\cite{KSVZ},
the KSVZ axions, and by Dine, Fischler and Srednicki as well as
Zhitnitskii~\cite{DFSZ}, the DFSZ axions. Both make axions
invisible by increasing $v_{PQ}$ (the larger $v_{PQ}$, the weaker
the axion-matter coupling) and obtain the axion through a singlet
scalar. In the KSVZ axion model the ordinary quarks and leptons
do not carry PQ charges, some heavy new quarks have to be
included which carry this quantum number. On the contrary, in the
DFSZ ordinary quarks and leptons do carry PQ charges, although
these fermions do not couple directly to the singlet, which
happens only at the loop level through interactions in the potential.

The possibility of an invisible axion makes the PQ approach even more
attractive since in this case the axion is a natural candidate for
explaining the existence of cold dark matter (CDM)~\cite{CDM}.
This is possible because the axion receives a tiny mass through
chiral anomaly, $m_a^2\sim \Lambda_{QCD}^4 /f_{PQ}^2$, amounting
to a mass of ${\cal O}(10^{-5})$~eV. However it is not easy, in
general, to obtain the required PQ symmetry in a natural way,
most models have to impose it from the beginning, weakening
such a solution to the strong-CP problem. That
is the reason we concentrate here in a class of models where the
symmetry would arise automatically, namely a particular version
of the $SU_C(3)\otimes SU_L(3)\otimes U_Y(1)$  model (3-3-1 for
short)~\cite{ppf,footpp,early,versionI,versionII}.

In 3-3-1 models the anomaly cancellation requires a minimal of
three families (or a multiple of three in larger versions).
Besides, there is a bunch of new particles and interactions which
make these models phenomenologically rich and attractive as an
alternative to the SM. If we assume that in the realm of
intermediate energy there are no exotic leptons, then the 3-3-1
symmetry allows for only two possible gauge models for the strong
and electroweak interactions, which will be referred as version I and
version II.

In the most popular one, version I~\cite{ppf}, the triplet of
leptons is composed of $(\nu_L \,, l_L \,,l^c_R)^T$, it contains
exotic quarks with electric charge $4/3$  and $5/3$, and doubly
charged bilepton gauge boson, $U^{\pm \pm}$, which prompts rare
lepton decays. It also implies an upper bound on the Weinberg
angle, $\sin(\theta_W)<1/4$. Version II is the 3-3-1 model with
right-handed neutrinos~\cite{footpp}. In it the triplet of
leptons is constituted by $(\nu_L \,, l_L \,, \nu^c_R)^T$. Its
bilepton gauge boson is neutral and their exotic quarks carry
usual charges, $1/3$ and $2/3$~\cite{f1}.

The physical properties of these models were investigated in
several works and their different aspects became
evident~\cite{versionI,versionII}. Among these differences it is
noticeable that version I requires a minimal of three triplets
and one sextet of scalars in order to generate the masses for all
fermions and gauge bosons while version II does the same job with
only three triplets.

Their shared aspects include the naturalness of massive neutrinos,
with the difference that in version I neutrinos are Majorana-type,
while in version II they are Dirac-type. Besides, from their
structure these models dispose of enough constraints upon the
$U(1)_N$ quantum numbers leading to the correct pattern for
electric charge quantization~\cite{pires}. Another of these
aspects is that also the PQ symmetry and the leptonic symmetry
can emerge naturally in both versions~\cite{pal,axion331,majoron}.

Since the version II of 3-3-1 was observed to possess the PQ
symmetry with a smaller content~\cite{majoron}, although in that
context the axion was of the Weinberg-Wilczek kind, we decided to
chose this more economical model and investigate the possibility
of obtaining an invisible axion by including only one extra scalar
singlet field in the model. The presence of CDM candidates in
version II of 3-3-1 was recently addressed~\cite{longlan}, but
here we wish to have the axion playing such a role. There is a
crucial issue that has to be addressed when trying to stick with
a CDM singlet axion though. It concerns the fact that gravitation
induces dangerous effective terms in the Lagrangian, explicitly
breaking any global symmetry of the theory. In particular,
focusing on $U_{PQ}(1)$, this breaking implies a huge
contribution to the axion mass. There remains the question
whether an appropriate mechanism exists in order to avoid such
terms, stabilizing the axion. Fortunately, the annoying terms can
be conveniently suppressed by the presence of suitable discrete
symmetries. Moreover, it was noticed in late eighties by Kraus and
Wilczek~\cite{kw}, that a local continuous symmetry at high
energies manifests at low energies as discrete (local) symmetries
which, differently from global ones, are expected to be respected
by gravity. This means that the needed discrete symmetries can
arise in a rather natural way if we assume some underlying local
continuous symmetry.

Discrete gauge symmetries have been used to stabilize the axion
in a model with extra-dimensions by Kamionkowsky {\it et
al.}~\cite{kamion} more than ten years ago. It was also pointed
out that large discrete symmetries can naturally arise in the
context of string theories~\cite{dine}. Also, in an attempt to
prevent $B-L$ violation in a class of supersymmetric standard
model, large discrete symmetries were imposed, implying an
automatic PQ symmetry, stabilized against quantum gravity
effects~\cite{lukas}. For what we are concerned, it was noticed
in Ref.~\cite{axion331} that 3-3-1 models possess a large enough
number of fields to accommodate large discrete symmetries, $Z_N$.
And the larger $N$ is, the higher are the number of suppressed
unwanted terms in the Lagrangian. In order to obtain a $Z_{13}$
symmetry, the authors in Ref.~\cite{axion331} added some extra
fermion fields to the model, resulting in an automatic PQ
symmetry and the axion protected under gravitational mass
corrections. This constitutes an additional motivation for
considering these 3-3-1 models to obtain the invisible axion and
solve the strong-CP problem.

This work is divided as follows. We first introduce the model in
section~\ref{sec1}. In section~\ref{sec2} we impose a
$Z_{11}\otimes Z_2$ symmetry, associating the appropriate charges
for the fields and obtain that the resulting Lagrangian is
invariant under $U_{PQ}(1)$, identifying the correct PQ charges.
This is done within the same spirit as that presented in
Ref.~\cite{axion331,axionsm}, assigning charges under a discrete
symmetry group to the fields at hand and observing that a PQ
symmetry emerges automatically if a $Z_2$ is also imposed. We
will see that in this case, also lepton number symmetry arises
naturally. In section~\ref{sec3}, we analyze the symmetry
breaking pattern of the model, recognizing the axion and its
couplings. We finally present the conclusions in
section~\ref{sec4}.

\section{The model }
\label{sec1}

Our investigation on this work relies on the version II of the
3-3-1 models~\cite{versionII}. Its left-handed lepton content
comes in the fundamental representation of the $SU(3)_L$,
composing the following triplet,
\bea f^a_L = \left (
\begin{array}{c}
\nu^a_L \\
e^a_L \\
(\nu^{c}_R)^a
\end{array}
\right )\sim(1\,,\,3\,,\,-1/3)\,, \eea
and the right-handed leptons are singlets,
\be e_{aR}\,\sim(1,1,-1)\,, \ee
with $a=1,2,3$ representing the three known generations. We are
indicating the transformation under 3-3-1 after the similarity
sign, ``$\sim$''. Differently from version I, right-handed
neutrinos are already present instead of exotic leptons.

In the quark sector, one generation of left-handed fields comes in
the triplet fundamental representation of $SU(3)_L$ and the other
two compose an anti-triplet with the following content,
\bea &&Q_{iL} = \left (
\begin{array}{c}
d_{iL} \\
-u_{iL} \\
d^{\prime}_{iL}
\end{array}
\right )\sim(3\,,\,\bar{3}\,,\,0)\,,\,\,\,Q_{3L} = \left (
\begin{array}{c}
u_{3L} \\
d_{3L} \\
u^{\prime}_{3L}
\end{array}
\right )\sim(3\,,\,3\,,\,1/3)\,, \label{quarks1} \eea
and the right-handed fields,
\bea
&&u_{iR}\,\sim(3,1,2/3)\,,\,d_{iR}\,\sim(3,1,-1/3)\,,\, d^{\prime}_{iR}\,\sim(3,1,-1/3)\nonumber \\
&&u_{3R}\,\sim(3,1,2/3)\,,\,d_{3R}\,\sim(3,1,-1/3)\,,\,u^{\prime}_{3R}\,\sim(3,1,2/3),
\label{quarks2} \eea
where $j=1,2$ represent different generations. The primed quarks
are the exotic ones but with the usual electric charges.

In order to generate the masses for the gauge bosons and fermions,
the model requires only  three triplets of scalars, namely,
\bea \chi = \left (
\begin{array}{c}
\chi^0 \\
\chi^{-} \\
\chi^{\prime 0}
\end{array}
\right )\sim(1\,,\,3\,,\,-1/3) \,,\,\,\, \eta = \left (
\begin{array}{c}
\eta^0 \\
\eta^- \\
\eta^{\prime 0}
\end{array}
\right )\sim(1\,,\,3\,,\,-1/3)\,,\,\,\, \rho = \left (
\begin{array}{c}
\rho^+ \\
\rho^0 \\
\rho^{\prime +}
\end{array}
\right )\sim(1\,,\,3\,,\,2/3)\,. \label{chieta}\eea
%
%

With these scalars and matter fields we can write the following
Yukawa interactions~\cite{f2},
\bea {\cal L}^Y&=& G_1\bar{Q}_{3L}u^{\prime}_{3R} \chi
+G_2^{ij}\bar{Q}_{iL}d^{\prime}_{jR}\chi^* +
G_3^{3a}\bar{Q}_{3L}u_{aR}\eta +G_4^{ia}\bar{Q}_{iL}d_{aR}\eta^*
\nonumber \\
&&+G_5^{3a}\bar{Q}_{3L}d_{aR}\rho
+G_6^{ia}\bar{Q}_{iL}u_{aR}\rho^*
+h_{ab}\bar{f}_{aL}e_{bR}\rho
+h^{\prime}_{ab}\epsilon^{ijk}(\bar{f}_{aL})_i(f_{bL})^c_j(\rho^*)_k
+\mbox{H.c.}. \label{yukintera} \eea

After the breaking of the 3-3-1 symmetry the vector gauge bosons
$W^{\pm}$,  $V^{\pm}$ , $U^0$ and $U^{0 \dagger}$ interact with
matter as follows~\cite{f3},
\bea {\cal L}^{CI}= -\frac{g}{\sqrt{2}}&&\left[
\bar{\nu}^a_{L}\gamma^\mu e^a_{L}W^+_\mu
+(\bar{\nu}^c_{R})^a\gamma^\mu e^a_{L}V^+_\mu+\bar{\nu}^a_{L}
\gamma^\mu (\nu^c_{R})^a U_\mu^0+
\bar{u}^a_{L}\gamma^\mu d^a_{L}W^+_\mu \right.\nonumber \\
&&\left.+(\bar{u}^{\prime}_{3L} \gamma^\mu
d_{3L}+\bar{u}_{iL}\gamma^\mu d^{\prime}_{iL})V^+_\mu +
(\bar{u}_{3L}\gamma^\mu
u^{\prime}_{3L}-\bar{d}^{\prime}_{iL}\gamma^\mu d_{iL})U^0_\mu
\right]+\mbox{H.c.}\,. \label{CC} \eea

It is through these Lagrangian interactions, ${\cal L}^Y$ and
${\cal L}^{CI}$, that we can recognize particles that carry
lepton number L such as total lepton number is conserved at this
level. From these interactions we have
\be {\mbox L}(V^+\,,\, u^{\prime}_{3} \,,\, \eta^{\prime
0}\,,\,\rho^{\prime +})=-2 \,,\,\,\,\,\,\,\,\,\,\,{\mbox L}(U^0
\,,\,d^\prime_{i} \,,\, \chi^ 0\,,\, \chi^-)=+2.
\label{leptonnumber} \ee

Notice that the new quarks, $u^{\prime}_3$ and $d^{\prime}_i$ are
leptoquarks once they carry lepton and baryon numbers; $V^{\pm}$
are charged vector bileptons while $U^0$ and $U^{0 \dagger}$ are
neutral vector bileptons. We have also charged scalar bileptons
and two neutral scalar bileptons. These last ones would be
important in studying spontaneous breaking of lepton number, if
the associated global symmetry is conserved by the potential,
leading to the so called majoron, as discussed in
Ref.~\cite{majoron}.

We include also an additional singlet scalar field, $\phi\sim
(1,1,0)$, in order to complete the spectrum, allowing for the
desired discrete symmetry which will enable us to get an axion
protected under large gravitational contribution to its mass.

Finally, we can write the most general, renormalizable and gauge
invariant, potential for this model. We divide it in two pieces,
one hermitean, $V_H$, and one non-hermitean, $V_{NH}$, which can
be written as,
\bea V_H &=& \mu_\phi^2 \phi^2 + \mu_\chi^2 \chi^2
+\mu_\eta^2\eta^2 +\mu_\rho^2\rho^2+\lambda_1\chi^4
+\lambda_2\eta^4
+\lambda_3\rho^4+
\lambda_4(\chi^{\dagger}\chi)(\eta^{\dagger}\eta)
+\lambda_5(\chi^{\dagger}\chi)(\rho^{\dagger}\rho) \nonumber \\
&&+\lambda_6
(\eta^{\dagger}\eta)(\rho^{\dagger}\rho)+
\lambda_7(\chi^{\dagger}\eta)(\eta^{\dagger}\chi)
+\lambda_8(\chi^{\dagger}\rho)(\rho^{\dagger}\chi)+\lambda_9
(\eta^{\dagger}\rho)(\rho^{\dagger}\eta)+
\lambda_{10} (\phi\phi^*)^2
\nonumber \\
&& +\lambda_{11}(\phi\phi^*)(\chi^{\dagger}\chi)
+\lambda_{12}(\phi\phi^*)(\rho^{\dagger}\rho) +
\lambda_{13}(\phi\phi^*)(\eta^{\dagger}\eta)\,, \label{VH} \eea
and
\bea V_{NH} &=& \mu_{\chi\eta}^2\chi^\dagger\eta +
f_1\chi^\dagger\eta\phi + f_2\chi^\dagger\eta\phi^* +
\lambda_{14}(\chi^\dagger\eta)^2
+ \lambda_{15}\chi^\dagger\eta\phi\phi+
\lambda_{16}\chi^\dagger\eta\phi^*\phi +
\lambda_{17}\chi^\dagger\eta\phi^*\phi^*
\nonumber \\ &&
+ \frac{1}{\sqrt{2}}\epsilon^{ijk}\left(f_3\eta_i\rho_j\chi_k
+f_4\eta_i\eta_j\rho_k + f_5\chi_i\chi_j\rho_k\right)
+ \epsilon^{ijk}\left(\lambda_{18}\eta_i\rho_j\chi_k +
\lambda_{19}\eta_i\eta_j\rho_k +
\lambda_{20}\chi_i\chi_j\rho_k\right)\phi
\nonumber \\ &&
+ \epsilon^{ijk}\left(\lambda_{21}\eta_i\rho_j\chi_k +
\lambda_{22}\eta_i\eta_j\rho_k +
\lambda_{23}\chi_i\chi_j\rho_k\right)\phi^*
+ \lambda_{24}(\chi^\dagger\rho)(\rho^\dagger\eta)
+\lambda_{25}(\chi^\dagger\eta)(\eta^\dagger\eta)
\nonumber \\
&&+\lambda_{26}(\chi^\dagger\eta)(\rho^\dagger\rho)
+\lambda_{27}(\chi^\dagger\eta)(\chi^{\dagger}\chi)\,+ H.c.\,.
\label{VNH} \eea

With this at hand we have all the necessary ingredients to
associate a discrete symmetry, $Z_{11}$, to the model. This will
allow us to eliminate several terms in the non-hermitean potential
Eq.~(\ref{VNH}) and verify that we need only an additional $Z_2$
symmetry to have PQ symmetry naturally, assigning the appropriate
PQ charges to fermions and scalars.

\section{$\mathbf{Z_{11}}$ and PQ symmetries}
\label{sec2}

A discrete symmetry $Z_N$ can naturally be accommodate when the
theory has enough number of fields in its spectrum. It was
observed that this is the case for the SM when some scalar
multiplets and right handed neutrinos are added~\cite{axionsm},
or for the minimal 3-3-1 model when only right handed neutrinos
need to be included~\cite{axion331}. It was obtained that a
$Z_{13}$ local symmetry could be imposed in this way, leading to
a natural PQ symmetry. We remark that this idea was first pursued
by Lazarides {\it et al.} in the context of models embedded in
superstring theories~\cite{laza}.


Here we are going to apply such idea to the version of 3-3-1
model presented in section~\ref{sec1}, that has right handed
neutrinos in its fundamental representation. It was observed
that an axion might be a natural outcome when a $Z_2$ symmetry was
imposed in this model~\cite{axion331}. Although this axion is of the
Weinberg-Wilczek kind, thus phenomenologically discarded~\cite{PecceiHomma},
if we consider the enlarged spectrum with a singlet scalar,
$\phi$, the axion can be a mixing of this field with other
scalars in the model, with its major component being the
pseudo-scalar part of the $\phi$ field. Then, a discrete symmetry
can be imposed allowing for an axion also protected under
gravitational mass corrections.

To proceed in this way we first assign the $Z_N$ charges to all
independent fields, and check for additional symmetries appearing
after eliminating forbidden terms under $Z_N$. It will turn out
that a chiral $U(1)$ symmetry arise, and we will see that it is
possible to identify it with PQ symmetry. It would be interesting
to have a $Z_{13}$ symmetry so as to obtain a PQ scale in its
upper limit, $v_{PQ}\sim 10^{12}$GeV. Although the model disposes
of 14 independent multiplets, it is not possible to accommodate a
symmetry greater than $Z_{12}$, because the Yukawa interactions in
Eq.~(\ref{yukintera}) imply some constraints over the allowed
$Z_N$ charges. It is clear that $N=12$ is the value of $N$ that
allows for a maximal protection of the axion under gravitational
effects in this model. However, besides the seemingly difficulty
of avoiding to repeat the phases of the multiplets, the singlet
$\phi$ would have to acquire a very specific phase since twelve is
not a prime number. In other words, any even phase would make the
transformation to belong to a smaller discrete symmetry,
jeopardizing our intent of suppressing some high order operators
involving $\phi$ products. For this reason the largest discrete
symmetry we can use is $Z_{11}$, which allows any phase to
$\phi$, except the trivial one.

The effective operators responsible for the gravitational mass
contribution are of the form $\phi^{n}/M_{Pl}^{n-4}$. A $Z_N$
symmetry automatically suppress terms of this kind till some
$n=N-1$. The main surviving term contributing to the axion mass is
the one with $n=N$. It is true that with $Z_{11}$ the axion is
protected only for energy scales not bigger than $\langle\phi
\rangle\simeq 10^{10}$~GeV. Nevertheless, this is not a threat
for the model since we still have values for the $\theta$ angle
and axion mass (gravitationally induced)~\cite{pecceinucl},
\bea M_a^{Grav}&\simeq &\sqrt{\frac{\langle\phi
\rangle^{N-2}}{M_{Pl}^{N-4}}}\simeq 10^{-12}\,\mbox{eV}\simeq
10^{-7}m_a\,, \nonumber \\ \nonumber \\
\theta_{eff}&\simeq &\frac{\langle\phi
\rangle^{N}}{M_{Pl}^{N-4}\Lambda_{QCD}^4}\simeq 10^{-19}\,,
\label{massateta} \eea
where we have used $M_{Pl}\simeq 10^{19}$~GeV and
$\Lambda_{QCD}\simeq 300$~MeV, and $m_a\simeq 10^{-5}$~eV is the
instanton induced axion mass. These values are consistent with
astrophysical and experimental bounds (see PDG~\cite{PDG1}). If we
had taken $\langle\phi \rangle\simeq 10^{11}$~GeV, the axion
would still be protected under gravitation, but the $\theta$
value would be on the threshold of its bound
$\theta_{eff}\lesssim 10^{-9}$. So we can have a valid solution
to the strong-CP problem for $Z_{11}$ for scales $\langle\phi
\rangle \,\lesssim \,10^{10}$~GeV in this version of 3-3-1. In
order to seek for this solution let us proceed further by first
assigning the correct $Z_{11}$ charges to the fields. Defining
$\omega_k \equiv e^{2\pi i\frac{k}{11}}\,,\,\{k=0,\pm 1,..., \pm
5\}$ the $Z_{11}$ transformations are given by:
\bea \phi\,\,\,\,\,&\rightarrow &\,\,\,\,\,
\omega_1\phi\,,\,\,\,\,\,\,\,\,\,\,\,\,\,\,\,\,\,\,\,\,\,\,\,\,\,f_{aL}\,\,\,\,\,\,\,
\rightarrow\,\,\,\,\,\,\,\omega_1^{-1}f_{aL}\,,
\nonumber \\
\rho\,\,\,\,\,&\rightarrow &\,\,\,\,\,
\omega_2\rho\,,\,\,\,\,\,\,\,\,\,\,\,\,\,\,\,\,\,\,\,\,\,\,\,\,\,d_{aR}\,\,\,\,\,\,\,
\rightarrow\,\,\,\,\,\,\,\omega_2^{-1}d_{aR}\,,
\nonumber \\
\chi\,\,\,\,\,&\rightarrow &\,\,\,\,\,
\omega_3\chi\,,\,\,\,\,\,\,\,\,(e_R ,u_{3R}^\prime)
\,\,\,\,\,\,\,\,\rightarrow\,\,\,\,\,\,\,\omega_3^{-1}(e_R
,u_{3R}^\prime)\,,
\nonumber \\
Q_{iL}\,\,\,\,\,&\rightarrow &\,\,\,\,\, \omega_4
Q_{iL}\,,\,\,\,\,\,\,\,\,\,\,\,\,\,\,\,\,\,\,\,
d_{iR}^\prime\,\,\,\,\,\,\,\rightarrow\,\,\,\,\,\,\,\omega_4^{-1}d_{iR}^\prime\,,
\nonumber \\
\eta\,\,\,\,\,&\rightarrow &\,\,\,\,\,
\omega_5\eta\,,\,\,\,\,\,\,\,\,\,\,\,\,\,\,\,\,\,\,\,\,\,\,\,\,\,u_{aR}\,\,\,\,\,\,\,
\rightarrow\,\,\,\,\,\,\,\omega_5^{-1}u_{aR}\,,
\nonumber \\
Q_{3L}\,\,\,\,\,&\rightarrow &\,\,\,\,\, \omega_0 Q_{3L} \,.
\label{z11cargas} \eea
At this point it is possible to go back to the potential,
Eq.~(\ref{VNH}), and note that this symmetry eliminates all
non-hermitean terms except three, namely,
$\chi^\dagger\eta\phi^*\phi^*,\,\,\eta\rho\chi\phi,\,\,
\eta\eta\rho\phi^*$.

If, besides the $Z_{11}$ symmetry we impose a $Z_2$ symmetry that
acts as,
\be (\phi\,,\chi\,,d_R^\prime\,,u_{3R}^\prime )
\,\,\,\,\rightarrow\,\,\,\,
-(\phi\,,\chi\,,d_R^\prime\,,u_{3R}^\prime ) \,, \label{z2} \ee
with the remaining fields transforming trivially, the only term
which remains in the non-hermitean potential is the
$\eta\rho\chi\phi$. It should be noted that the Yukawa
interactions in Eq.~(\ref{yukintera}) do not allow for terms
which interchange $\chi\leftrightarrow\eta$, since they do not
respect $Z_{11}\otimes Z_2$ given by Eqs.~(\ref{z11cargas}) and
(\ref{z2}).

We have the stage settled to see that an automatic PQ symmetry arise
in the model. To achieve this conclusion we start by
assigning the PQ quantum numbers such that quarks of opposite
chiralities have opposite charges, yielding chiral quarks under
$U_{PQ}(1)$ transformation,
\bea u_{aL}&\rightarrow & e^{-i\alpha X_u}u_{aL}\,,\,\,\,\,
u_{aR}\,\,\,\rightarrow\,\,\, e^{i\alpha X_u}u_{aR}\,, \nonumber \\
u_{3L}^\prime &\rightarrow & e^{-i\alpha
X_u^\prime}u_{3L}^\prime\,,\,\,\,\,
u_{3R}^\prime\,\,\,\rightarrow\,\,\,
e^{i\alpha X_u^\prime}u_{3R}^\prime\,, \nonumber \\
d_{aL}&\rightarrow & e^{-i\alpha X_d}d_{aL}\,,\,\,\,\,
d_{aR}\,\,\,\rightarrow\,\,\, e^{i\alpha X_d}d_{aR}\,, \nonumber \\
d_{iL}^\prime &\rightarrow & e^{-i\alpha
X_d^\prime}d_{iL}^\prime\,,\,\,\,\,\,\,
d_{iR}^\prime\,\,\,\rightarrow\,\,\, e^{i\alpha
X_d^\prime}d_{iR}^\prime\,. \label{pqcargasQ} \eea

For the leptons we can define their PQ charges by,
\bea e_{aL}&\rightarrow & e^{i\alpha X_e}e_{aL}\,,\,\,\,\,
e_{aR}\,\,\,\rightarrow\,\,\, e^{i\alpha X_{eR}}e_{aR}\,, \nonumber \\
\nu_{aL} &\rightarrow & e^{i\alpha X_\nu}\nu_{aL}\,,\,\,\,\,
\nu_{aR}\,\,\,\rightarrow\,\,\, e^{i\alpha X_{\nu R}}\nu_{aR}\,.
\label{pqcargasL} \eea

With these assignments and taking the Yukawa interactions in
Eq.~(\ref{yukintera}) into account, as well as the non-hermitean
terms $\eta\rho\chi\phi$, we easily see that the PQ charges for
the scalars are constrained and imply the following relations:
\be
X_d=-X_u\,,\,\,\,\,X_{d^\prime}=-X_{u^\prime}\,,\,\,\,\,X_\nu=X_{eR}
\,,\,\,\,\,X_e=X_{\nu R}\,. \label{pqvinculos} \ee
We can make the further choice $X_d=X_{d^\prime}$, leading to
\be X_d=X_{d^\prime}=-X_u=-X_{u^\prime}=-X_e=X_{eR}=X_\nu=-X_{\nu
R}\,, \label{pqc} \ee
implying that the PQ symmetry is chiral for the leptons too,
and the scalars transform as,
\bea \phi &\rightarrow & e^{-2i\alpha
X_d}\phi\,,\,\,\,\,\,\,\,\,\,\,\,\,\,\,
\eta^0\,\,\,\,\rightarrow\,\,\, e^{2i\alpha X_d}\eta^0 \nonumber \\
\eta^-&\rightarrow &
\eta^-\,,\,\,\,\,\,\,\,\,\,\,\,\,\,\,\,\,\,\,\,\,\,\,\,\,\,\,\,\,\,
\eta^{\prime 0}\,\,\,\rightarrow\,\,\, e^{2i\alpha X_d}\eta^{\prime 0} \nonumber \\
\rho^+&\rightarrow &
\rho^+\,,\,\,\,\,\,\,\,\,\,\,\,\,\,\,\,\,\,\,\,\,\,\,\,\,\,\,\,\,\,\,
\rho^{0}\,\,\,\,\rightarrow\,\,\, e^{-2i\alpha X_d}\rho^{0} \nonumber \\
\rho^{\prime +}&\rightarrow & \rho^{\prime
+}\,,\,\,\,\,\,\,\,\,\,\,\,\,\,\,\,\,\,\,\,\,\,\,\,\,\,\,\,\,\,
\chi^{0}\,\,\,\rightarrow\,\,\, e^{2i\alpha X_d}\chi^{0} \nonumber \\
\chi^{-}&\rightarrow &
\chi^{-}\,,\,\,\,\,\,\,\,\,\,\,\,\,\,\,\,\,\,\,\,\,\,\,\,\,\,\,\,\,\,
\chi^{\prime 0}\,\,\,\rightarrow\,\,\, e^{2i\alpha X_d} \chi^{\prime 0} \nonumber \\
\label{pqcargasS} \eea

Now it is transparent that the whole Lagrangian of the model is
$U_{PQ}(1)$ invariant and the strong-CP problem can be solved in
the context of this model.
The strong-CP violation angle is given
by the sum over the quarks PQ charges, which translates to
\be \theta \longrightarrow \theta \pm 2\alpha X_d\,. \label{teta}
\ee
This result is possible in this version of 3-3-1 because the PQ
charges of the exotic quarks, $d^\prime$ and $u_3^\prime$ do not
cancel exactly for the case of interest here, $X_u = -X_d$.
Moreover, the model is particularly attractive in the sense it does not
present the domain wall problem~\cite{domainwall}. This means that
there is no discrete subset of PQ symmetry that leaves the axion
potential invariant, i.e., $\nexists Z_N\subset U_{PQ}(1)$ such
that $V_{axion}(\theta )$ is invariant. This is similar to what
happens in the 3-3-1 version discussed in Ref.~\cite{axion331},
although there right-handed neutrinos had to be added to the
model besides the singlet scalar.

It is remarkable that under $Z_{11}\otimes Z_2$, not only the PQ
chiral symmetry is automatic but the lepton number symmetry also
appears naturally in the model, once the possibly non-conserving
lepton number terms present in the potential completely
disappeared. In this sense, discrete symmetries originating at
some high energy scale seems to be enough to generate the desired
global symmetries we need at lower energies.

We finally write the most general potential invariant under 3-3-1
and $Z_{11}\otimes Z_2$ (or $U_{PQ}(1)$ and Lepton number)
symmetries ,
\bea V(\eta,\rho,\chi)&=& V_{H}+
\lambda_{\phi}\epsilon^{ijk}\eta_i\rho_j\chi_k \phi+ H.c.\,,
\label{firstpot} \eea
where $V_{H}$ is given in Eq.~(\ref{VH}).

In the next section we are going to use this potential to
recognize the axion, the Goldstone boson originating from the
breaking of the PQ symmetry, and verify that it is constituted
mostly of the singlet $\phi$.

\section{Spontaneously broken PQ symmetry}
\label{sec3}

The potential given in the previous section, Eq.~(\ref{firstpot}),
allows us to obtain the mass eigenstates for the scalars, so we
can identify the Goldstones which are absorbed by the massive
gauge bosons and extract the axion in terms of the interaction
eigenstates. To accomplish this, let us consider that only $
\chi^{\prime 0} $, $ \rho^0 $, $\eta^0$ and $\phi$ develop a vacuum
expectation value (VEV) and expand such fields around their VEV's
in the standard way,
\bea \chi^{\prime 0} &=&  \frac{1}{\sqrt{2}} (v_{\chi^{\prime}}
+R_{\chi^{\prime}} +iI_{\chi^{\prime}})\,,\,\,\,\,\,\,\, \eta^ 0=
\frac{1}{\sqrt{2}} (v_{\eta} +R_{\eta} +iI_{\eta})\,, \nonumber \\
\rho^0 &=& \frac{1}{\sqrt{2}} (v_{\rho} +R_{\rho}
+iI_{\rho})\,,\,\,\,\,\,\,\,\,\,\,\,\,\,\,\, \phi=
\frac{1}{\sqrt{2}} (v_{\phi} +R_{\phi} +iI_{\phi})\,.
\label{vacua} \eea
With such expansion, the next step is to get the constraints that
lead to the minimum of the potential,
\bea &&\mu^2_\chi +\lambda_1 v^2_{\chi^{\prime}} +
\frac{\lambda_4}{2}v^2_\eta +
\frac{\lambda_5}{2}v^2_\rho+\frac{\lambda_{11}}{2}v_\phi^2
+\frac{A}{v_{\chi^\prime}^2} =0,\nonumber \\
&&\mu^2_\eta +\lambda_2v^2_\eta + \frac{\lambda_4}{2}
v^2_{\chi^{\prime}} +\frac{\lambda_6}{2}v^2_\rho
+\frac{\lambda_{13}}{2} v^2_{\phi}+\frac{A}{v_\eta^2} =0,\nonumber \\
&&\mu^2_\rho +\lambda_3 v^2_\rho + \frac{\lambda_5}{2}
v^2_{\chi^{\prime}}
+\frac{\lambda_6}{2}v^2_\eta+\frac{\lambda_{12}}{2}
v^2_{\phi}+\frac{A}{v_\rho^2} =0,
\nonumber \\
&&\mu^2_\phi +\lambda_{10}v^2_{\phi} +
\frac{\lambda_{11}}{2}v^2_{\chi^{\prime}}+\frac{\lambda_{12}}{2}v^2_\rho+
\frac{\lambda_{13}}{2}v^2_\eta +\frac{A}{v_\phi^2}
=0\,,\label{mincondI} \eea
where we have defined $A\equiv\lambda_{\phi}v_\eta v_\rho
v_{\chi^{\prime}}v_\phi$. Substituting the expansion in
Eq.~(\ref{vacua}) in the potential Eq.~(\ref{firstpot}) and using
the constraints above, we get the following mass matrix,
$M_R^2(R_\chi\,,R_{\eta^{\prime}})$, for the real scalars in the
basis, $(R_\chi\,,R_{\eta^{\prime}})$,
\bea\left(\begin{array}{cc} \frac{\lambda_7v^2_\eta}{4}-\frac{A}{2
v_{\chi^\prime}^2}
& \frac{\lambda_7v_{\chi^{\prime}}v_\eta }{4}-\frac{A}{2 v_{\chi^{\prime}}v_\eta} \\
\frac{\lambda_7v_{\chi^{\prime}}v_\eta }{4}-\frac{A}{2
v_{\chi^{\prime}}v_\eta} & \frac{\lambda_7v^2_{\chi^{\prime}}}{4}
-\frac{A}{2 v_\eta^2}
\end{array}
\right)\,, \label{matrixR1} \eea
and the mass matrix, $M_R^2(R_{\chi^{\prime}}\,,R_\eta\,,R_\rho\,
,R_\phi )$, in the basis $(R_{\chi^{\prime}}\,,R_\eta\,,R_\rho\,
,R_\phi )$,
\bea
\left(\begin{array}{cccc}  \lambda_1
v^2_{\chi^{\prime}}-\frac{A}{2 v_{\chi^\prime}^2} &
\frac{\lambda_4 v_{\chi^{\prime}}v_\eta}{2}+\frac{A}{2 v_\eta
v_{\chi^\prime}} & \frac{\lambda_5 v_{\chi^{\prime}}v_\rho}{2}
+\frac{A}{2 v_\rho v_{\chi^\prime}} & \frac{A}{2 v_\phi v_{\chi^\prime}} \\
\frac{\lambda_4 v_{\chi^{\prime}}v_\eta}{2}+\frac{A}{2 v_\eta
v_{\chi^\prime}} & \lambda_2 v^2_\eta -\frac{A}{2 v_\eta^2} &
\frac{\lambda_6 v_\eta
v_\rho}{2} +\frac{A}{2 v_\rho v_\eta} & \frac{A}{2 v_\eta v_\phi}  \\
\frac{\lambda_5 v_{\chi^{\prime}}v_\rho}{2} +\frac{A}{2 v_\rho
v_{\chi^\prime}} & \frac{\lambda_6 v_\eta v_\rho}{2} +\frac{A}{2
v_\rho v_\eta} & \lambda_3v^2_\rho-\frac{A}{2 v_\rho^2} &
\frac{A}{2 v_\rho v_\phi} \\
\frac{A}{2 v_\phi v_{\chi^\prime}} & \frac{A}{2 v_\eta v_\phi} &
\frac{A}{2 v_\rho v_\phi} & \lambda_{10} v_\phi^2 -\frac{A}{2
v_\phi^2}
\end{array}
\right). \label{matrixR2}
\eea

These bases are not coupled, that is the reason we have two
squared mass matrices. From the first matrix,
Eq.~(\ref{matrixR1}), after diagonalization it is easy to
recognize the following massless scalar in it,
\be R_G=\frac{1}{ \sqrt{v^2_\eta +v^2_{\chi^{\prime}}}}(v_\eta
R_{\eta^{\prime}} - v_{\chi^{\prime}}R_\chi)\,.
\label{realgoldstone} \ee
The other real scalar mass eigenstate is orthogonal to this one
and those coming from the diagonalization of matrix
Eq.~(\ref{matrixR2}), which are a little more intricate but
fortunately we do not need them for our purpose.

Regarding the pseudo-scalars, similarly to the real scalars, we
obtain the following mass matrix,
$M_I^2(I_\chi\,,I_{\eta^{\prime}})$ in the basis $(I_\chi
\,,I_{\eta^{\prime}})$,
\be \left(\begin{array}{cc} \frac{\lambda_7v^2_\eta}{4} -
\frac{A}{2 v_{\chi^\prime}^2} &
-\frac{\lambda_7v_{\chi^{\prime}}v_\eta}{4} +\frac{A}{2 v_\eta v_{\chi^\prime}} \\
-\frac{\lambda_7v_{\chi^{\prime}}v_\eta}{4} +\frac{A}{2 v_\eta
v_{\chi^\prime}} & \frac{\lambda_7v^2_{\chi^{\prime}}}{4} -
\frac{A}{2 v_\eta^2}
\end{array}
\right)\,, \label{matrixI1} \ee
and the mass matrix,
$M_I^2(I_{\chi^\prime}\,,I_\eta\,,I_\rho\,,I_\phi )$, in the basis
$(I_{\chi^\prime}\,,I_\eta\,,I_\rho\,,I_\phi )$,
\be - \frac{A}{2}\left(\begin{array}{cccc}  \frac{1}{
v_{\chi^\prime}^2} & \frac{1}{ v_\eta v_{\chi^\prime}} &
\frac{1}{ v_\rho v_{\chi^\prime}}
& \frac{1}{ v_{\chi^\prime}v_\phi} \\
\frac{1}{ v_\eta v_{\chi^\prime}} &  \frac{1}{ v_\eta^2} &
\frac{1}{ v_\eta v_\rho} & \frac{1}{ v_\eta v_\phi}\\
\frac{1}{ v_\rho v_{\chi^\prime}} & \frac{1}{ v_\eta v_\rho}
& \frac{1}{ v_\rho^2} & \frac{1}{ v_\eta v_\rho} \\
\frac{1}{ v_{\chi^\prime}v_\phi} & \frac{1}{ v_\eta v_\phi} &
\frac{1}{ v_\eta v_\rho} & \frac{1}{ v_\phi^2}
\end{array}
\right)\,. \label{matrixI2} \ee
From these matrices we can easily obtain the Goldstone bosons and
identify the axion as the one whose main component is in the
$\phi$ direction. The Goldstones and the pseudo-Goldstones are
listed below,
\bea a &=&
\frac{1}{\sqrt{1+\frac{v_{\chi^\prime}^2}{v_\phi^2}}}\left(I_\phi
-\frac{v_{\chi^\prime}}{v_\phi}I_{\chi^\prime}\right)\,,
\nonumber \\
G_1 &=&
\sqrt{\frac{\xi}{v_\phi^2v_{\chi^\prime}^2+
\xi}}\left(I_\rho
-\frac{v_\phi^2v_{\chi^\prime}}{\xi}I_{\chi^{\prime}}
 -\frac{v_\phi
v_{\chi^\prime}^2}{\xi}I_\phi\right)\,,
\nonumber \\
G_2 &=&
\frac{1}{\sqrt{1+\frac{v_{\chi^\prime}^2}{v_\eta^2}}}\left(
I_{\eta^\prime}+ \frac{v_{\chi^\prime}}{v_\eta}I_\chi \right)\,,
\nonumber \\
G_3 &=&
\sqrt{\frac{v_\eta^2(v_\phi^2v_{\chi^\prime}^2+\xi )}
{v_\rho^2v_\phi^2v_{\chi^\prime}^2+v_\eta^2(v_\phi^2v_{\chi^\prime}^2+
\xi )}}\left(I_\eta
-\frac{v_\rho^2v_\phi^2v_{\chi^\prime}}{v_\eta(v_\phi^2v_{\chi^\prime}^2+
\xi )}I_{\chi^\prime}
-\frac{v_\rho
v_\phi^2v_{\chi^\prime}^2}{v_\eta(v_\phi^2v_{\chi^\prime}^2+
\xi )}I_\rho -\frac{v_\rho^2v_\phi
v_{\chi^\prime}^2}{v_\eta(v_\phi^2v_{\chi^\prime}^2+
\xi )}I_\phi\right)\,, \nonumber
\\
PS_1 &=& \frac{1}{\sqrt{1+\frac{v_\phi^2}{v_\eta^2}+
\frac{v_\phi^2}{v_\rho^2}+\frac{v_\phi^2}{v_{\chi^\prime}^2}}}\left(
I_\phi + \frac{v_\phi}{v_{\chi^\prime}}I_{\chi^\prime} +
\frac{v_\phi}{v_\eta}I_\eta +\frac{v_\phi}{v_\rho}I_\rho\right)\,,
\nonumber \\
PS_2 &=&
\frac{1}{\sqrt{1+\frac{v_\eta^2}{v_{\chi^\prime}^2}}}\left(
I_{\eta^\prime} - \frac{v_\eta}{v_{\chi^\prime}} I_\chi\right)\,,
\label{goldstones}\eea
where we have defined $\xi \equiv
v_\rho^2(v_\phi^2+v_{\chi^\prime}^2)$. In the above equation, the
axion is identified as $a$, and the remaining three Goldstones,
$G_1$, $G_2$ and $G_3$, together with the Goldstone in
Eq.~(\ref{realgoldstone}) are those eaten by the 4 neutral gauge
bosons of the model. The last two linear combinations of the
interaction states in Eq.~(\ref{goldstones}), $PS_1$ and $PS_2$,
are the massive eigenstates or pseudo-Goldstones. The important
point that can be extract from these results is that our axion
has a small component of $I_{\chi^\prime}$. Since, $v_\phi\simeq
10^{10}$GeV and $v_{\chi^\prime}\simeq 10^3$GeV, this component
is very suppressed and, as expected, our axion is invisible being
almost exclusively the imaginary part of the singlet $\phi$.
Besides, since $I_{\chi^\prime}$ couples only to the exotic
quarks, our axion is very different even from that obtained in
the version of 3-3-1 in Ref.~\cite{axion331}, which does couple to
neutrinos at tree level. Its coupling can be easily obtained
after rotating the mass eigenstates, Eq.~(\ref{goldstones}), in
terms of the interaction eigenstates, and it translates into the
following Lagrangian term,
\be {\cal L}_{aq^\prime q^\prime}=
\frac{-iv_{\chi^\prime}}{\sqrt{2(v_\phi^2+v_{\chi^\prime}^2)}}
\left[G_1\bar{u}_{3L}^\prime u_{3R}^\prime -G^{ij}_2
\bar{d}_{iL}^\prime d_{jR}^\prime \right]a + \mbox{H.c.}\,,
\label{qqa}\ee
which are very weak for $G_1,G^{ij}_2\sim$ of the order of unit,
since $v_\phi\gg v_{\chi^\prime}$. The pseudo-Goldstones,
$PS_1$ and $PS_2$, are more strongly coupled
to fermions and,
differently from the axion, also couple to ordinary matter. This
leads us to conclude that, in this model, the only candidate for
cold dark matter is the axion. We could check if the real massive
scalars could fit for this role also, but a rough numerical
approximation just confirms that they behave as their partners,
as we could expect.

We also checked the coupling of our axion with photons. It is
defined through the effective Lagrangian term,
\be {\cal
L}_{a\gamma\gamma}=\frac{\bar{c}_{a\gamma\gamma}}{32\pi^2}
\frac{a(x)}{v_{PQ}}\tilde{F}_{\mu\nu}F^{\mu\nu}\,.
\label{axionfotons}\ee
In the present model only exotic quarks participate in the loop
leading to the above anomaly term, which leads to
\be \bar{c}_{a\gamma\gamma}=-\frac{2}{3v_{PQ}}\sum_{q^\prime}
X_{q^\prime} Q_{q^\prime}^2 = \frac{4}{9}\simeq 0.44\,.
\label{couplingagg} \ee
This value is very similar to those obtained in different models
present in literature and can be used to make the relevant
computations involving axion in astrophysical processes.

\section{Summary and conclusions}
\label{sec4}

We studied the consequences of discrete symmetries in the version
of 3-3-1 model with right-handed neutrinos. One of the main
points in this work is that global symmetries appear
automatically as consequence of such discrete symmetries in this
model.  It turned out that, when the model has a $Z_{11}\otimes
Z_2$ symmetry, the whole Lagrangian is invariant under $U(1)$
transformations and also total lepton number is conserved at the
classical level. It is remarkable that this happens in this more
economical version of the model by adding only one singlet
scalar, no other fields are necessary, which makes it a suitable
model for implementing the strong-CP problem solution.

We then recognized the global symmetry identifying it with a
chiral PQ symmetry, $U_{PQ}(1)$. In general, solutions to the
strong-CP problem through PQ mechanism lead to the domain wall
formation, which is a threatening feature to Cosmology, but
fortunately model dependent. In this version of 3-3-1 model this
problem is absent due to the fact that we chose a relation among
PQ charges, namely $X_d=X_d^\prime$, which avoid this situation.
Nevertheless, we have to remark that other possibilities would be
allowed if we had not imposed a $Z_2$ symmetry. In this case, we
could have let $X_d^\prime$ free and, working with the
non-hermitean potential terms, look for the relations among the
PQ charges that would keep the $U_{PQ}(1)$ invariance. Such a
relation exists and is given by $3X_d^\prime = -X_d$, which is
consistent with PQ invariance for all three $Z_{11}$ invariant
non-hermitean terms, $\chi^\dagger\eta\phi^*\phi^*$,
$\eta\rho\chi\phi$ and $\eta\eta\rho\phi^*$. In this situation we
would have to address the formation of domain walls, and for this
reason our previous choice seemed physically more appealing.

We proceeded with the spontaneous breaking of $U_{PQ}(1)$ and studied its
consequences, obtaining the axion and
showing that it is mainly constituted of the singlet. It was shown
that it interacts
with exotic quarks only, through a very suppressed
coupling. Hence, our axion is an invisible one. However,
since global symmetries are not stable against gravity effects,
our axion could be in danger, receiving large mass corrections and
losing its appealing as CDM candidate. This was circumvented by
assuming that $Z_{11}$ is a subgroup of an underlying gauge group
at some high energy scale. Annoying gravity induced terms
contributing to the axion mass, are conveniently suppressed thanks
to the local $Z_{11}$, stabilizing the axion.
Although, larger discrete symmetries
would lead to a better stabilization of the axion against gravity, as
well as a $\theta_{eff}$ safer from experimental constraints, we saw
that is still possible to have a $Z_{11}$ leading to small
mass corrections
and $\theta_{eff}$ below the bound $\theta_{eff}< 10^{-9}$ if
$v_{\phi}\equiv\langle\phi\rangle\simeq 10^{10}$~GeV. Besides,
the fact that 11 is a prime number allows $\phi$ to acquire any
charge under $Z_{11}$ except the trivial one. This implies no
need of assigning a very specific charge to $\phi$ in order to
avoid restriction to smaller subsets of $Z_N$ which would not
lead to axion stabilization.

Finally, there is a point that has to be highlighted when
considering the 3-3-1 version with right handed neutrinos. In the
form it was presented here the model generates arbitrary masses
for two neutrinos only, which can be deduced from the
anti-symmetry of the Yukawa coupling $h_{ab}\prime$ in
Eq.~(\ref{yukintera}). This leads to a anti-symmetric neutrino
mass matrix, implying a zero eigenvalue and two degenerate ones.
Although the massiveness of neutrinos is not the issue here, it
would be nice to have a model that at least produces an
appropriate arbitrary mass pattern for all fermions of the
theory. To accomplish this we have to devise some way of
eliminating such mass degeneracy. We can suggest two ways of doing
that. One of them makes no enlargement of the spectrum and seems
the preferable one, dealing only with the vacuum of the theory.
In Ref.~\cite{majoron}, the breaking of leptonic symmetry could
be achieved only through a non-conserving PQ symmetry term,
namely the $\eta\rho\chi$, when $\eta^\prime$ acquires a
non-vanishing VEV. However, we have seen that our approach allows
for an equivalent term which could lead to lepton number
violation too, which is $\eta\rho\chi\phi$, with the difference
that in this case PQ symmetry is still conserved. This term
allows for a Majorana neutrino mass through loop corrections,
making possible a mass matrix which is arbitrary enough to
accommodate non-degenerate and non-zero neutrino masses. Another
way out could be traced by including a singlet neutrino in the
spectrum. Such neutrino carries the exact quantum numbers to
provide the required invariance under $Z_{11}\otimes Z_2$ and
$U_{PQ}(1)$ and leave only non-degenerate massive neutrinos in
the theory. This second possibility sounds appealing since not
many fields can be introduced without jeopardizing the desired
discrete and global symmetries here studied. Whatever nature's
choice, both would be adequate to fit in our approach.

Summarizing, we obtained automatic $U_{PQ}(1)$ and lepton number
symmetries by imposing a local discrete symmetry in an economic
version of 3-3-1 model with right-handed neutrinos. We got an
invisible axion stabilized under gravitational mass corrections,
absent of domain wall problem, solving the strong-CP problem and
constituting a strong candidate to CDM. This results are a
remarkable achievement of our work, considering that the only
additional ingredient we have used was the inclusion of a singlet
scalar in the model.

\section*{Acknowledgments}
We would like to thank Vicente Pleitez for useful discussions and
suggestions.
This research was partially supported by Funda\c{c}\~ao de Amparo
\`a Pesquisa do Estado de S\~ao Paulo (FAPESP) (AGD,PSRS) and by
Conselho Nacional de Desenvolvimento Cient\'{\i}fico e
Tecnol\'ogico (CNPq) (CASP).


\begin{thebibliography}{99}
\bibitem{f1} These new exotic quarks
of both versions carry baryon and lepton numbers as well, and are
fairly recognized as leptoquarks.
\bibitem{f2} Observe that $\chi$ and
$\eta$ transforming in the same way should have similar couplings
with matter. However, the interactions here presented are those
which generate the desired mass pattern for the fermions. Later
we will see that they are the only ones that will survive under
the discrete symmetries of the model.
\bibitem{f3} We are omitting the neutral current
interactions here because throughout this work we will be dealing
only with interactions which might provide some global quantum
number violation.
\end{thebibliography}
\end{document}